\documentclass[%
 reprint,
 amsmath,amssymb,
 aps,
prb,
floatfix,
]{revtex4-2}

\usepackage{graphicx}
\usepackage{dcolumn}
\usepackage{bm}

    \usepackage[title]{appendix}
    \usepackage{comment}
    \usepackage[dvipsnames]{xcolor}
    \definecolor{darkblue}{RGB}{0,0,50.2}
    \definecolor{steelblue}{RGB}{70,130,180}
    \definecolor{mygreen}{RGB}{28,172,0}
    \definecolor{mylilas}{RGB}{170,55,241}
    \usepackage{hyperref}
    \hypersetup{
        colorlinks=true,
        linkcolor=steelblue,
        filecolor=magenta,      
        urlcolor=steelblue,
        citecolor=mylilas,
        pdftitle={Superconducting Stiffness and Coherence Length of FeSe$_{0.5}$Te$_{0.5}$ Measured in Zero-Applied Field},
        pdfpagemode=FullScreen,
        }
    \usepackage{epsfig}
    \usepackage{float}
    \usepackage{cancel}
    \usepackage{mathtools}

\begin{document}
    
    
\title{Superconducting Stiffness and Coherence Length of FeSe$_{0.5}$Te$_{0.5}$ Measured in Zero-Applied Field}

\author{Amotz Peri}
    \email{amotzpery@gmail.com}
\author{Itay Mangel}%
\author{Amit Keren}%
    \email{keren@physics.technion.ac.il}
\affiliation{%
Department of Physics, Technion-Israel Institute of Technology, Haifa, 3200003, Israel}%


\date{\today}

\begin{abstract}
Superconducting stiffness $\rho_s$ and coherence length $\xi$ are usually determined by measuring the penetration depth $\lambda$ of a magnetic field and the upper critical field $H_{c2}$ of a superconductor (SC), respectively. However, in magnetic SC, e.g. some of the iron-based, this could lead to erroneous results since the internal field could be very different from the applied one. To overcome this problem in Fe$_{1+y}$Se$_x$Te$_{1-x}$ with $x \sim 0.5$ and $y \sim 0$ (FST), we measure both quantities with the Stiffnessometer technique. In this technique, one applies a rotor-free vector potential $\textbf{A}$ to a superconducting ring and measures the current density $\textbf{j}$ via the ring's magnetic moment $\textbf{m}$. $\rho_s$ and $\xi$ are determined from London’s equation $\textbf{j}=-\rho_s\textbf{A}$ and its range of validity. This method is particularly accurate at temperatures close to the critical temperature $T_c$. We find weaker $\rho_s$ and longer $\xi$ than existing literature reports, and critical exponents which agree better with expectations based on the Ginzburg–Landau theory.
\end{abstract}

\maketitle




\section{\label{sec:introduction}INTRODUCTION}

The highest $T_c$ measured in bulk Iron-based superconductors (IBSs), in ambient pressure, is 56~K \cite{Wang2008Thorium-dopinginducedGdThxFeAsO}, higher than some cuprates, \textit{e.g.} optinaly doped La$_{2-x}$Sr$_{x}$CuO$_4$. Consequently, they have been at the forefront of research in the solid-state community. Out of all IBSs, the crystalline structure of the FeSe is the simplest. By partially replacing Se with Te atoms, the critical temperature increases up to 15~K, obtained at $x=0.45$ $y\simeq 0$ in the formula Fe$_{1+y}$Se$_{x}$Te$_{1-x}$. As summarized by Kreisel {\it et al.}~\cite{Kreisel2020OnCousins}, the material also possesses surprising properties such as highly anisotropic electronic properties (nematic effects) and evidence for topologically non-trivial bands and superconductivity. In light of these properties, it is important to characterize Fe$_{1+y}$Se$_{x}$Te$_{1-x}$ as accurately as possible. Here we focus on the $x\sim 0.5$ $y \sim 0$ variant (FST), which is available as bulk crystal.

Bulk DC superconducting properties, such as the stiffness $\rho_s$, were measured in this crystal by transverse field muon spin rotation ($\mu$SR) \cite{Biswas2010Muon-spin-spectroscopy0.5Se0.5,Bendele2010AnisotropicTe0.5}. AC measurements were done by RF tunnel diode \cite{Serafin2010AnisotropicTe0.5,Kim2010London} and cavity perturbation \cite{Takahashi2011AnomalousCrystals,kurokawa2021relationship} techniques. The Coherence length of FST with $x=0.45$ was determined by vortex size $\xi$ using a scanning tunneling microscope (STM) \cite{Wang2018EvidenceSuperconductor} and resistivity measurements \cite{Shruti2015AnisotropyFeTe0.55Se0.45}. The Cooper pair size $\xi_0$ was evaluated with angle-resolved photoemission spectroscopy (ARPES) \cite{Chiu2020ScalableSuperconductors}. However, due to the presence of Fe in the structure and residual magnetism, the field dependent measurements might not provide a clear insight into the superconducting properties since the applied field interacts with a magnetic moment in addition to the superconducting currents. In this work, we measure DC superconducting properties in a zero-applied field to avoid contamination from magnetism.

The superconducting stiffness $\rho_s$ is defined via the gauge-invariant relation between the current density $\textbf{j}$, the total vector potential $\textbf{A}_{\text{tot}}$ from all sources, and the complex order parameter $\Psi(\textbf{r})=\psi(\textbf{r})\text{e}^{\text{i} \phi(\textbf{r})}$ with $\psi(\textbf{r}) > 0$, according to 
\begin{equation}
\textbf{j}=-\rho_s(\textbf{A}_{\text{tot}}-\frac{\Phi_0}{2\pi}\bm{\nabla}\phi)\,,
\label{eq:London gauge invariant}
\end{equation}
where $\Phi_0=2\pi\hbar/e^*$ is the superconducting flux quanta, 
\begin{equation}
\rho_s=\frac{\psi^2e^{*2}}{m^*}\,,
\label{eq:stiffness}
\end{equation}
is the stiffness, $e^*$ and $m^*$ are the carrier's charge and mass, respectively. For anisotropic stiffness see Ref.~\cite{Kapon2019Phase:viewpoint}. When cooling the superconductor (SC) with $\textbf{A}_{\text{tot}}=0$ in the London gauge, minimum kinetic energy requires $\bm{\nabla}\phi=0$. According to the second Josephson relation, $\phi$ can only change by dissipating energy. Thus, Eq.~\ref{eq:London gauge invariant} becomes the London equation
\begin{equation}
\textbf{j}=-\rho_s\textbf{A}_{\text{tot}}\,.
\label{eq:London}
\end{equation}

This relation holds as long $\bm{\nabla}\phi$ does not change. The stiffness, in turn, is related to the penetration depth via
\begin{equation}
\rho_s=\frac{1}{\mu_0\lambda^2}\,.
\label{eq:stiffness and lambda}
\end{equation}

However, every superconductor has a critical current density $j_c$ determined by the penetration depth $\lambda$ and coherence length $\xi$. When $\textbf{A}_\text{tot}$ exceeds a certain value, it is worthwhile for the SC to change $\bm{\nabla}\phi$ so as to keep $j$ below $j_c$ everywhere in the SC. According to the Josephson equation, dynamic changes in $\phi$ lead to voltage, which, when combined with current, result in power and energy dissipation in the process. When this happens, the relation between $\textbf{j}$ and $\textbf{A}_{\text{tot}}$ is no longer linear and the system's rigidity breaks. We used these properties to measure both $\rho_s$ and $\xi$ as a function of temperature in FST.

\section{\label{sec:experiment setup}EXPERIMENT SETUP}

\begin{figure}[h]
    \centering
    \includegraphics[width=0.9\linewidth]{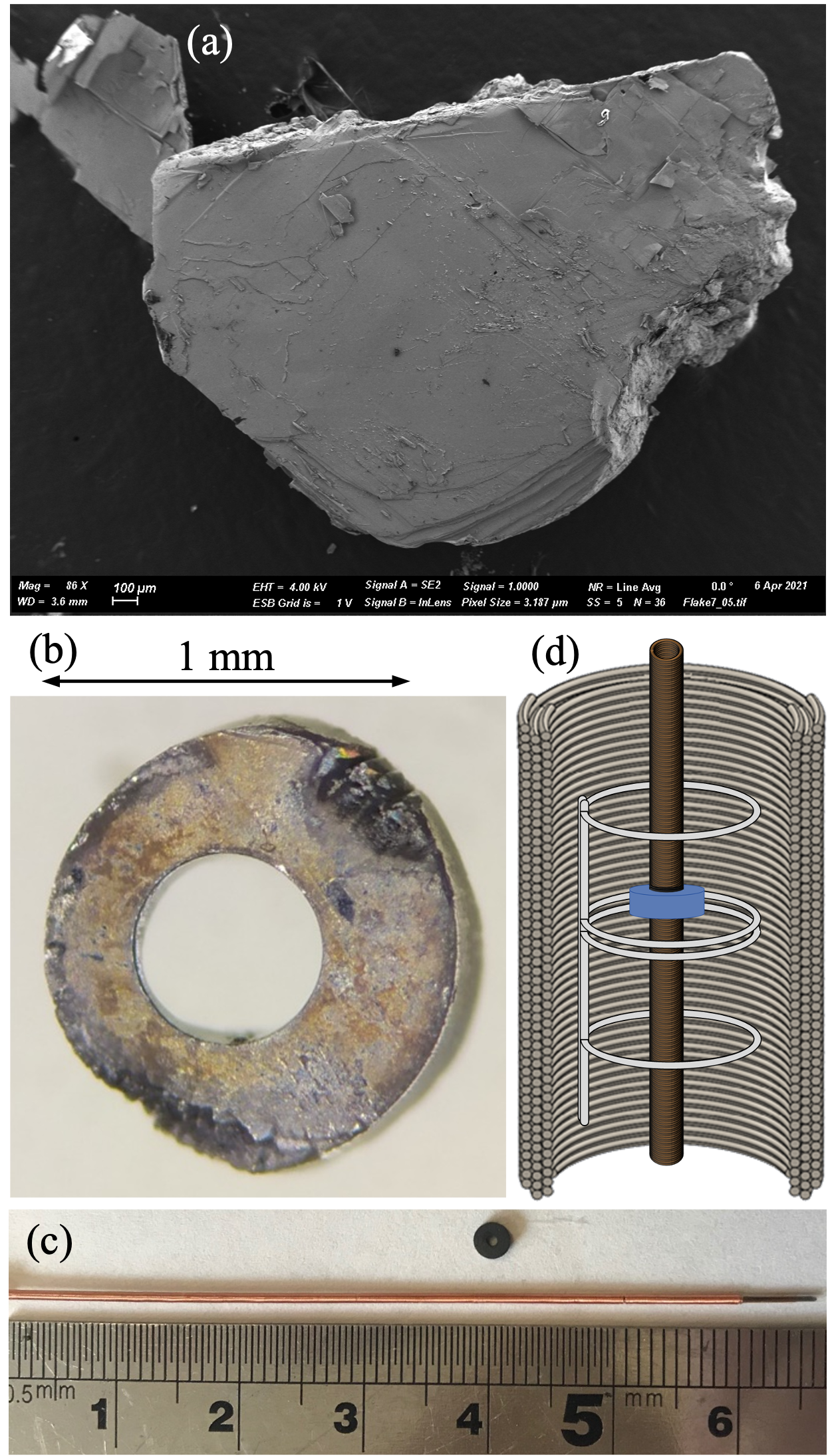}
    \caption{ {\bf Experimental setup}: (a) A scanning electron microscope image of a single crystal of FST, from which the ring was cut out. (b) A microscopic image of the Fe$_{1+y}$Se$_{0.5}$Te$_{0.5}$ ring. The sample is not uniform. The minimal height, inner, and minimal outer radii are $h=0.10$~mm, $r_\text{in}=0.26$~mm, and $r_\text{out}=0.50$~mm respectively. (c) A copper excitation coil and a superconducting ring beside it. The coil has a length of 60~mm, an outer diameter of 0.25~mm, and 9300 turns in two layers. (d) The ring and excitation coil assembly is moving rigidly relative to a gradiometer, connected to a SQUID system (not shown), and surrounded by a main coil for field zeroing or field-dependent measurement. The SQUID, gradiometer, and main coil are part of a QD-MPMS3 system.}
    \label{fig:setup}
\end{figure}

The experiment is assembled from a ring-shaped SC cut out of a single crystal, shown in Fig.~\ref{fig:setup}(a), with a femtosecond laser. The ring is presented in panel (b). Since FST is brittle, the ring is not perfect. But, as we argue below (in Sec.~\ref{subsec:stiffness model}), the smallest outer radius and height count for our analysis. The ring is pierced by a long excitation coil (EC). These parts are shown in panel (c). The excitation coil, ring, and second-order gradiometer are surrounded by a main coil, as in panel (d). The main coil is used to zero the field to less than 0.1~$\mu$T on the ring, and for field-dependent measurement. Details of the dimensions of the different parts are given in the figure's caption. EC current $I_\text{ec}$ generates a vector potential $\textbf{A}_{\text{ec}}$ on the ring, nominally without a magnetic field $H$. This vector potential is responsible for persistent rotational current in the superconducting ring. This rotational current produces its own vector potential $\textbf{A}_{\text{sc}}$ and a magnetic moment. The vector potential in Eq.~\ref{eq:London gauge invariant} is $\textbf{A}_{\text{tot}}=\textbf{A}_{\text{ec}}+\textbf{A}_{\text{sc}}$. The sample's magnetic moment $m$ is detected by vibrating the ring with the EC rigidly relative to the gradiometer. This mode is called vibrating sample magnetometer (VSM) mode. It utilizes a lock-in amplifier to measure the SQUID output voltage at twice the vibration frequency. This output is proportional to the magnetic flux of a sample through the gradiometer, namely, the vector potential of the sample. It could also be represented by the magnetic moment of the sample.

The gradiometer is composed of two outer loops wound clockwise, and inner two loops wound anticlockwise, see Fig.~\ref{fig:setup}(d). In that way, we separate the magnetic signal generated by the sample from any other field uniform in space, even if it drifted in time. The gradiometer, main coil, and SQUID are part of QD-MPMS3 magnetometer. 

In principle, $\textbf{A}_{\text{ec}}$ does not change as the coil vibrates since there are no EC flux $\Phi_\text{ec}$ variations, and the pickup-loop signal is only due to $\textbf{A}_{\text{sc}}$. In practice, the small signal of the EC is reduced from the measurements as background, (see Sec.~\ref{subsec:stiffness}). The ring's vector potential at a pickup-loop radius $R_\text{pl}$ at $z=0$ is related to $\textbf{m}$ in the EC direction $\hat{z}$, by 
\begin{equation}
\textbf{A}_{\text{sc}}(r=R_\text{pl},z=0)=\frac{\mu_{0}}{4\pi}\frac{m}{R_{\text{pl}}^{2}}\hat{\varphi}\,.
\label{eq:SC VP to moment}
\end{equation}
where $\hat{\varphi}$ is the azimuth direction.


\section{\label{sec:measurements}Measurements}

\subsection{\label{subsec:stiffness} Stiffness and critical current}

\begin{figure}[h]
    \centering
    \includegraphics[width=\linewidth]{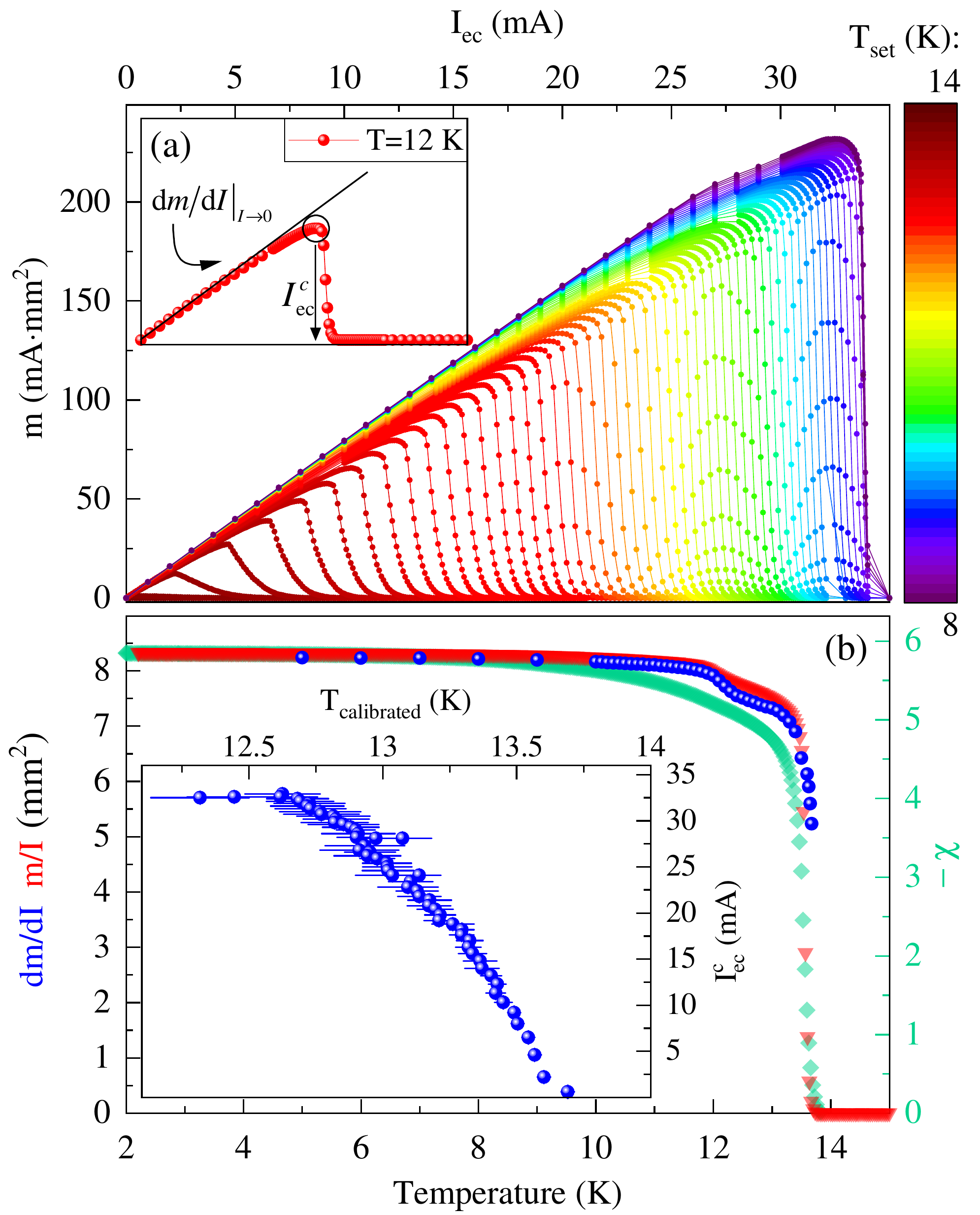}
    \caption{{\bf Data:} (a) Stiffness measurements. SC's magnetic moment vs. the current in the excitation coil at different temperatures, indicated by the colors. The inset is focused on measurement at 12~K. A linear relation is found for low currents. At some critical current value, the signal drops to zero. The blue circles in (b) depict the temperature dependence of the linear slope obtained at low currents (far from $I_\text{ec}^{c}$) in panel (a). (b) Critical temperature. SC's moment over current in the EC vs. the temperature (red down-triangles) as described in Sec.~\ref{subsec:stiffness}; measured susceptibility (with a minus sign) vs. the temperature (emerald diamonds) in MKS units in the presence of a magnetic field of $~1$~mT and without an excitation coil (according to Sec.~\ref{subsec:Tc}). Inset (b) shows the critical currents vs. the calibrated temperature (extracted from the breakpoints in panel (a)).}
    \label{fig:MvsI_and_dMdIvsT}
\end{figure}

We cool the system below $T_c$ with $I_\text{ec}=0$. After the temperature has stabilized, we gradually increase $I_\text{ec}$ while measuring the superconducting ring's magnetic moment. An example of a measurement at $T=12$~K is presented in Fig.~\ref{fig:MvsI_and_dMdIvsT}(a)-inset. A repetition of this process at different temperatures appears in panel (a). To isolate the superconducting signal we subtract the moment of the measurement with zero current, which is due to the ferromagnetic properties of FST and not its stiffness. In addition, we remove the current dependent of the signal above $T_c$. This signal is due to the EC's finite length and asymmetry.

Typical behavior in our measurements, for low currents, is a linear relation between the ring's moment and $I_\text{ec}$,  as expected (see Ref.~\cite{Gavish2021GinzburgLandauDevice} and Sec.~\ref{sec:mathematical model}). At some value of  $I_\text{ec}$, which defines the critical current $I_\text{ec}^{c}$, this relation breaks. Beyond the breakpoint, the magnetic signal drops sharply instead of the saturation behavior seen previously \cite{Mangel2020Stiffnessometer:Application,Keren2021StiffnessSuperconductors}. This drop is a result of two effects: I) heat produced by the copper EC, which leads to a temperature gradient between the ring and the thermometer. II) heat produced by energy dissipation as vortices enter the sample and $\phi$ changes dynamically. In fact, when the moment drops to zero, the ring has passed its critical temperature and stops being superconducting. A simple solution to the undesired effect (I) could have been to use a superconducting coil, but, the $T_c$ of FST is higher than any commercially available superconducting wire. Instead, we calibrated the temperature at the ring position using an open ring. The calibration is discussed in Appx.~\ref{Apx:T_calibration}. 

To extract the stiffness, we fit each $m(I_\text{ec})$ to a line in a temperature-dependent range due to the variation in the critical current. Such a fit is demonstrated in the inset of Fig.~\ref{fig:MvsI_and_dMdIvsT}(a). The slope represents $\text{d}m/\text{d}I$ in the limit $I_\text{ec}\rightarrow0$. The temperature dependence of the slopes appears as blue circles in Fig.~\ref{fig:MvsI_and_dMdIvsT}(b). The measurements do not cover all the temperature ranges up to $T_c$ since it becomes exceedingly difficult to define a linear region in the $m(I_\text{ec})$ data. At a temperature slightly below $T_c$, a knee appears in the temperature dependence of $\text{d}m/\text{d}I$.

The red down triangles in Fig.~\ref{fig:MvsI_and_dMdIvsT}(b) measure $m/I$ as a function of $T$. This is done by cooling with $I_\text{ec}=10$~mA, turning the current off and warming while measuring. At $T>13.45$~K, this current is above $I_\text{ec}^{c}$, and such a measurement cannot be used to extract the stiffness near $T_c$. On the other hand, such measurement can be carried out all the way to $T_c$. Interestingly, the knee is observed even with this constant current measurement. It is important to mention that the knee was detected in other FST rings as well. A detailed discussion on the knee, is given in Sec.~\ref{subsec: Knee Discussion}.

Finally, in Fig.~\ref{fig:MvsI_and_dMdIvsT}(b)-inset, we present $I_\text{ec}^{c}(T)$, corresponding to the moment's maximum, as a function of the calibrated temperature. The large error bars at the low temperatures range are due to the strong current in the coil, leading to a significant temperature gradient and uncertainty in the temperature calibration.

\subsection{\label{subsec:Tc} Susceptibility}
The emerald diamonds in Fig.~\ref{fig:MvsI_and_dMdIvsT}(b), depict the temperature dependence of the measured, zero-field cooled (ZFC) susceptibility $\chi=m/(HV_\text{ring})$, with a field of $\mu_0H=0.98$~mT parallel to the axial direction of the ring; $V$ stands for the ring's volume. The specific susceptibility is related to the measured one by 
\begin{equation}
    \chi=\frac{\chi_0}{1+D\chi_0}\,,
\end{equation}
where $D$ is the demagnetization factor, and $\chi_0$ is the specific susceptibility. For a ring with our geometry, the demagnetization factor equals $D=0.6$, and if we consider the inner radius of the ring $r_\text{in}\rightarrow0$, since in ZFC, it is hard for the field to penetrate the ring hole, $D=0.7$ \cite{Beleggia2009DemagnetizationShapes}. With these $D$ values (considering the effective volume of the ring in the latter case), we obtain, at $T\rightarrow 0$, $\chi_0=-1.30$, and $\chi_0=-1.15$, respectively.  $\chi_0=-1$ is excepted in the case that all of the ring's volume is superconducting. The extra 15\% or more in $\chi_0$ could be a result of the irregular shape of the ring. In any case, it indicates that the entire sample is superconducting. 

As for the temperature dependence of $\chi$, a sharp transition is observed towards the critical temperature in this measurement $T_c=13.82$~K, which indicates the quality of the material. Interestingly, in DC magnetization measurements, the knee is not observed.

\subsection{\label{subsec:Hysteresis_loop} Hysteresis}

\begin{figure}[h]
    \centering
    \includegraphics[width=\linewidth]{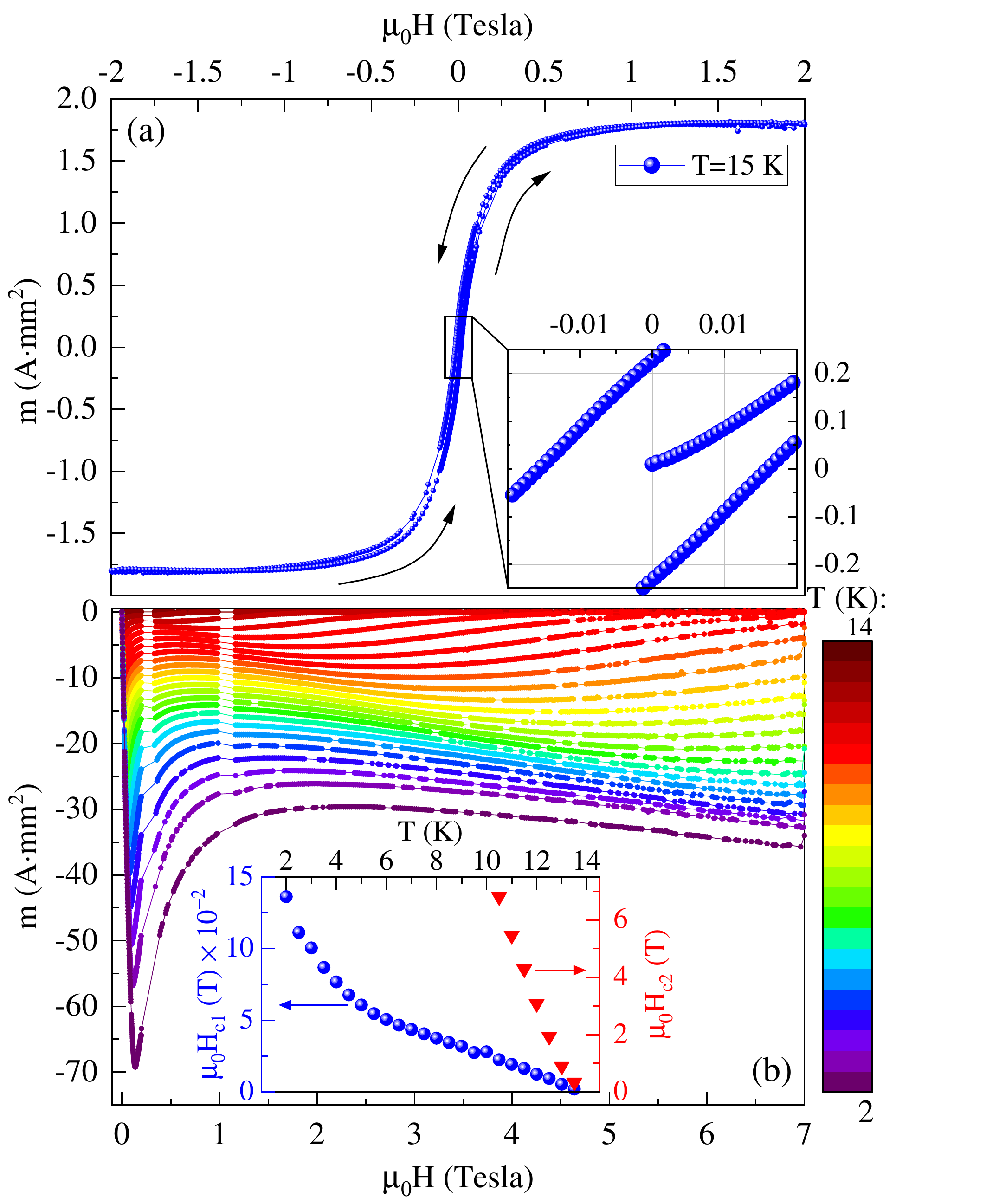}
    \caption{ {\bf Magnetic measurements}: (a) Magnetic hysteresis loop above the critical temperature. (b) $m(H)$ at different temperatures below $T_c$, as indicated by the colors. Inset: The temperature dependence of the critical fields $H_{c1}$ (blue circles) and $H_{c2}$ (red down-triangles) on the left and right Y-axis, respectively.}
    \label{fig:MvsH}
\end{figure}

To characterize the magnetic properties of the FST sample, we performed a magnetic hysteresis loop measurement, between $2$~T and $-2$~T, which is depicted in Fig.~\ref{fig:MvsH}(a). This measurement is done above the critical temperature, at $T=15$~K. The opening of a hysteresis loop is an indication of ferromagnetism.  Another sign is that the moment of the first point, at $H=0$, is different from zero. It might be difficult to notice this in the figure. However, this feature makes it possible to detect the sample without applying fields or currents above and below $T_c$, in contrast to non-magnetic materials. Additional properties that can be deduced from this measurement are the magnetization saturation, retentivity (remanence), and coercivity values: $m_\text{sat}=1.58$~A$\cdot$mm$^2$, $m_\text{remanence}=0.22$~A$\cdot$mm$^2$, and $\mu_0H_\text{coercivity}=0.0153$~T, respectively. Although this Ferromagnetism is sometimes ascribed to topological surface state \cite{PhysRevLett.130.046702} we analyze it as a bulk property. From the magnetization saturation and the magnetic moment of a free Fe ion $m_{\text{Fe}^{2+}}=5.4\mu_B$ or $m_{\text{Fe}^{3+}}=5.9\mu_B$, where $\mu_B$ is the Bohr magneton \cite{Weiss1976D.30.00.}, we can deduce that the fraction of the free iron ions per unit formula in the sample is $y=0.009$ or $y=0.008$, respectively. Wang {\it et al.}~\cite{Wang20220.5}, performed inelastic neutron scattering measurements of $\text{Fe}_{0.98}\text{Se}_{0.5}\text{Te}_{0.5}$ and claimed that $m_\text{Fe}=2.85$~$\mu_B$. The corresponding value for the iron fraction in our sample is $y=0.017$.

\subsection{\label{subsec:Hc} Critical magnetic fields}
The response of the superconducting ring to an applied magnetic field at different temperatures below $T_c$ is reflected in Fig.~\ref{fig:MvsH}(b). From that measurement, we extract the first and second critical fields, $H_{c1}$ and $H_{c2}$. $H_{c1}$ is defined by the maximum magnitude of the moment for each temperature. A second peak emerges at an intermediate field between $H_{c1}$ and $H_{c2}$, and is attributed to the role of twin boundaries \cite{Galluzzi2017EvidenceSuperconductor}. In principle, $H_{c2}$ is defined by the value of $H$ for which $m=0$ ~\cite{Tinkham2004IntroductionSuperconductivity}. However, it is not easy to determine $H_{c2}$ because of the asymptotic behavior of the moment. Therefore, we chose a criterion by which $H_{c2}$ is the field at which the moment is 10\% of the second peak magnitude. Below a temperature of $10$~K, $H_{c2}$ becomes higher than the maximum field available to us. $H_{c1}$ and $H_{c2}$ as a function of temperature are shown in the inset of panel (b).

\section{\label{sec:mathematical model} Analysis Model}

The analysis of Stiffnessometer data is described in details in Ref.~\cite{Gavish2021GinzburgLandauDevice} and is valid for systems with cylindrical symmetry. Here we provide only the major steps.

\subsection{\label{subsec:stiffness model}Stiffness}

In the low flux regime (low currents in the EC), the magnitude of the order parameter is constant almost all over the superconducting ring and zero outside \cite{Gavish2021GinzburgLandauDevice}. Substituting $\textbf{B}=\bm{\nabla}\times\textbf{A}$ and London equation into amperes law gives
\begin{equation}
    \bm{\nabla}\times\bm{\nabla}\times\textbf{A}_\text{sc}=-\mu_0\rho_s\textbf{A}_\text{tot}\,,
\end{equation}
since on the ring $\bm{\nabla} \times \textbf{A}_{\text{ec}}=0$. In the London gauge $\bm{\nabla}\times\bm{\nabla}\times\textbf{A}=-\nabla^2\textbf{A}$, and the vector potential outside an infinitely long coil is given by
\begin{equation}
    \textbf{A}_\text{ec}(r)=\Phi_{\text{ec}}/(2\pi r)\hat{\varphi}\,.
    \label{eq:EC VP to flux}
\end{equation}

With Eq.~\ref{eq:stiffness and lambda} we arrive at the partial differential equation (PDE)
\begin{equation}
    \nabla^2\textbf{A}_\text{sc}=\frac{1}{\lambda^2}\Big(\textbf{A}_\text{sc}+\frac{\Phi_{\text{ec}}}{2\pi r}\hat{\varphi}\Big)\,,
    \label{eq:PDE1}
\end{equation}
where $\lambda=\infty$ outside the SC. Normalizing the spatial variables and vector potential as follows
\begin{equation}
    \textbf{r}/R_\text{pl}\rightarrow\textbf{r}\;,\quad\textbf{A}_\text{sc}/\textbf{A}_\text{ec}(R_\text{pl})\rightarrow\textbf{A}\;,\quad\lambda/R_\text{pl}\rightarrow\lambda ,
    \label{eq:dimensionless transform}
\end{equation}
and using cylindrical coordinates where $\textbf{A}=A(r,z)\hat{\varphi}$, we end up with the following PDE
\begin{equation}
    \frac{\partial^2A}{\partial z^2}+\frac{\partial^2A}{\partial r^2}+\frac{1}{r}\frac{\partial A}{\partial r}-\frac{A}{r^2}=\frac{1}{\lambda^2}\Big(A+\frac{1}{r}\Big)\,.
    \label{eq:PDE dimensionless}
\end{equation}
We use the finite element-based FreeFem{\scriptsize ++} software~\cite{Hecht2012NewFreefem+} to solve this PDE for different values of $\lambda$ and the dimension of our FST ring, appearing in the caption of Fig.~\ref{fig:setup}. The equation is solved in a box such that $z\in[-L,\,L]$, and $r\in[0,\,8L]$ with $L=R_\text{pl}=8.5$~mm. Dirichlet boundary conditions are imposed.

As shown in Fig.~\ref{fig:setup}(b), the ring's outer radius is not uniform. However, the solution of the Ginzburg-Landau (GL) equations \cite{Gavish2021GinzburgLandauDevice} shows that in the low flux regime, the current flow in a layer of width $\lambda$ near the inner rim of the ring, so the system's symmetry is not severely compromised. When the flux through the ring is increased, the current layer retreats toward the outer rim. This retraction ends when the current layer reaches the outer rim. In our case, we assume that it happens at the shortest distance of the outer rim from the center. We use this distance as the outer radius in the PDE~\ref{eq:PDE dimensionless}. Nevertheless, our assumption has not been tested numerically and the impact of a non perfect ring on the result is not clear yet.

The red line in the inset of Fig.~\ref{fig:length_scale}(a) depicts the numerical solution of PDE~\ref{eq:PDE dimensionless}. The Y-axis is the normalized vector potential $A$ at the ring's height $z=0$, and the pickup-loop radial location $r=R_\text{pl}$. The X-axis is  $(R_\text{pl}/\lambda)^2$ on a logarithmic scale.
Normalizing Eq.~\ref{eq:SC VP to moment} by the vector potential of an infinite coil,
\begin{equation}
A_{\text{ec}}(R_\text{pl})=\frac{\mu_{0}nI_\text{ec}}{2R_\text{pl}}\sum_i r_{\text{ec},i}^2\,,
\label{eq:EC VP}
\end{equation}
where $n$ and $r_{\text{ec},i}$ are windings per unit length in one layer, and radius of the $i^\text{th}$ layer, respectively, we obtain the dimensionless vector potential
\begin{equation}
A(z=0,R_\text{pl})=\frac{g}{2\pi n R_\text{pl}\sum_ir_{\text{ec},i}^2}\cdot\frac{m}{I_\text{ec}}\,,
\label{eq:normalized VP}
\end{equation}
where $m$ is the SC's magnetic moment, and $g$ a geometrical constant on the order of unity.

In reality, the coil is not infinite, and, as a result of cutting and drilling, the ring is not perfect, see Fig.~\ref{fig:setup}(b). Therefore, the calibration constant $g$ is determined experimentally in two different methods: (1) We compare the saturated value of $A$ from the solution of PDE~\ref{eq:PDE dimensionless} (see red line in Fig.~\ref{fig:length_scale}(a)-inset) to the saturated value of $\text{d}m/\text{d}I$ [the lowest available temperature of the blue circles in Fig.~\ref{fig:MvsI_and_dMdIvsT}(b)]. This method cannot be used to determine $\lambda(T \rightarrow 0)$ since exactly this limit is used for the calibration.  Nevertheless, it gives one value for $g$; (2) We use a literature value of a low-temperature stiffness of similar material to predict $A$ with the PDE solution and compare it to our measured $\text{d}m/\text{d}I$ at the same temperature to extract a second value for $g$. For this work, the stiffness was taken from Ref.~\cite{Biswas2010Muon-spin-spectroscopy0.5Se0.5}. We found $g_1=0.5363$ and $g_2=0.5336$ in methods (1) and (2), respectively. We also applied the same calibration methods for a ring-shaped Niobium with similar dimensions and found  $g_1=g_2=0.68674(2)$ while using $\lambda(0)=39$~nm as the literature value for Niobium \cite{Varmazis1974InductiveDepth}. Although the two calibration methods give different values for the penetration depth at low temperatures, towards $T_c$, the values converge and almost coalesce, as we demonstrate shortly. In other words, the stiffness determined by the Stiffnessometer is not sensitive to the calibration method once $\text{d}m/\text{d}I$ is out of the saturation region.

\subsection{Coherence length}

In the low flux regime $\Phi_\text{ec}/\Phi_0 \ll r_{\text{in}}^{2}/\lambda\varepsilon$, and for $\lambda \ll r_\text{out}-r_\text{in}$ and $h$, where $h$ is the ring's height, deep inside the ring $A_\text{tot}=0$, hence $A_\text{sc}=-A_\text{ec}$. In other words, the applied flux is matched by the flux generated by the ring in the hole.  For $\Phi_\text{ec}/\Phi_0>r_\text{in}^2/\sqrt{8}\xi\lambda$, the current necessary to produce $A_\text{sc}$ at $r_\text{in}$ exceeds the local critical current \cite{Gavish2021GinzburgLandauDevice}. Then, it is energetically preferable for the order parameter magnitude to gradually diminish in the inner rim of the ring. Consequently, the superconducting ring hole effectively grows, and an effective inner radius $r_\text{eff}$ is established. 
At even higher flux, $r_\text{eff}$ approaches $r_\text{out}$, and the SC is no longer able to expel the applied flux, namely, to cancel $A_\text{ec}$. This happens at a critical flux \cite{Gavish2021GinzburgLandauDevice}
\begin{equation}
    \frac{\Phi_\text{c}}{\Phi_0} = \frac{r_\text{out}^2}{\sqrt{8}\xi\lambda}\,.
\label{eq:J fold}
\end{equation}

While the derivation of Eq.~\ref{eq:J fold} is in the limit $\xi \ll \lambda \ll r_\text{out}-r_\text{in} \ll h$, we believe it is valid for $\lambda \ll r_\text{out}-r_\text{in}$ and $\lambda \ll h$ separately. 

For $\Phi>\Phi_c$, vortices are expected to penetrate from the inner rim towards the outer one so that the SC's moment no longer grows with amplification of $I_\text{ec}$. These vortices are manifested in $\bm{\nabla}\phi$ variations.


\section{\label{sec:DA} Data analysis }

\begin{figure}[h]
    \centering
    \includegraphics[width=\linewidth]{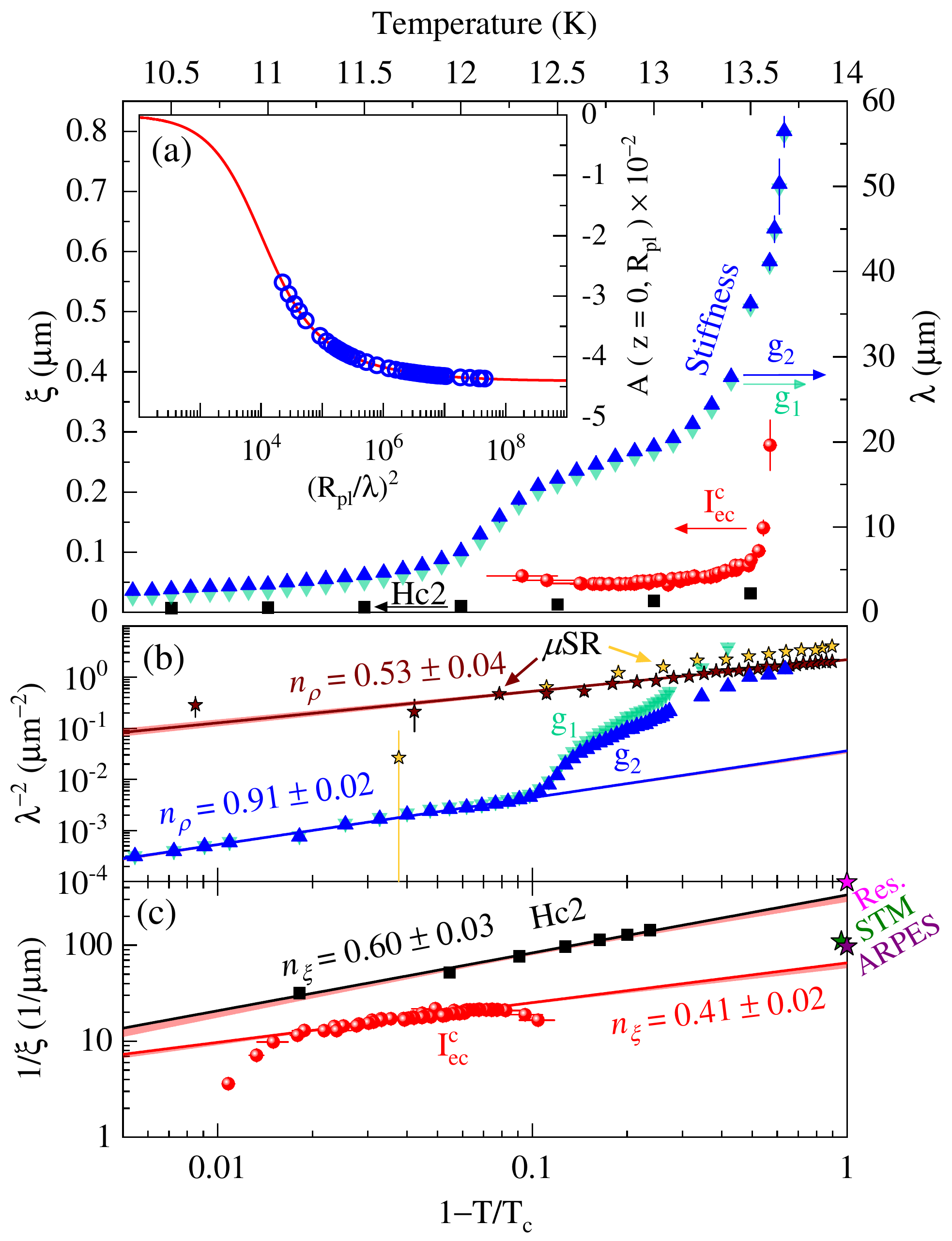}
    \caption{{\bf Penetration depth and coherence length.} (a) Right Y-axis, the penetration depth as a function of the temperature in blue and emerald triangles, for the different calibration methods. The left Y-axis shows the temperature dependence of the coherence length. The red-circles are taken from the critical current measurement in Fig.~\ref{fig:MvsI_and_dMdIvsT}(b)-inset through Eq.~\ref{eq:J fold} with the measured $\lambda$, and the black-squares are from the second critical field in Fig.~\ref{fig:MvsH}(b)-inset with Eq.~\ref{eq:Hc2 and xi}. Panels (b) and (c) are log-log plots of the stiffness $\lambda^{-2}$ and $1/\xi$ vs. $1-T/T_c$, respectively. The linear regression represents the critical exponents according to Eq.~\ref{eq:critical exponent of rho} and Eq.~\ref{eq:critical exponent of xi}, respectively. Earlier stiffness measurements using the $\mu$SR method have been added to (b) in brown \cite{Biswas2010Muon-spin-spectroscopy0.5Se0.5} and yellow \cite{Bendele2010AnisotropicTe0.5} stars. The same power law is fitted to this data. For comparison, we add to (c) in stars-shaped measurements of $1/\xi$ from the resistivity method \cite{Shruti2015AnisotropyFeTe0.55Se0.45} in magenta, ARPES \cite{Chiu2020ScalableSuperconductors} in purple, and STM \cite{Wang2018EvidenceSuperconductor} in green. }
    \label{fig:length_scale}
\end{figure}

Eq.~\ref{eq:normalized VP} relates the linear slope of the $m(I)$ measurements, shown by blue circles in Fig.~\ref{fig:MvsI_and_dMdIvsT}(b), to the numerical solution of the PDE.  The blue open circles in Fig.~\ref{fig:length_scale}(a)-inset represent the converted points using $g_2$. Each of those points belongs to a different temperature and gives a unique stiffness value. The temperature dependence of $\lambda$ is presented on a linear scale in  Fig.~\ref{fig:length_scale}(a), right Y-axis and of $\lambda^{-2}$ on a logarithmic scale in Fig.~\ref{fig:length_scale}(b), for the two different $g$ values. The difference between the two calibration methods is revealed in both sub-figures, but they are minute at $T\rightarrow T_c$. The linear regression towards the critical temperature on the logarithmic scale represents the critical exponent $n_\rho$, according to the power-law
\begin{equation}
    \rho\propto(1-T/T_c)^{n_\rho}\,,
    \label{eq:critical exponent of rho}
\end{equation}
with $n_\rho=0.91\pm0.02$. This relation describes the data well from the knee temperature $12.4$~K all the way to $T_c$. For comparison, the $\mu$SR measurements of $1/\lambda^2$ \cite{Biswas2010Muon-spin-spectroscopy0.5Se0.5,Bendele2010AnisotropicTe0.5} are also added to Fig.~\ref{fig:length_scale}(b), and their $n_\rho=0.53\pm0.04$. It should be pointed out that all techniques agree that $\lambda(T=0) \sim 0.5$~$\mu$m, but from tunnel diode technique $\lambda(T=0.9T_c) \sim 2$~$\mu$m \cite{Kim2010London}, which is longer than $\mu$SR but shorter than the Stiffnessometer.

Based on the stiffness and the critical current in the inset of Fig.~\ref{fig:MvsI_and_dMdIvsT}(b), we extract the coherence length using Eq.~\ref{eq:J fold} and the calculated flux in the coil. The results are depicted on a linear scale in Fig.~\ref{fig:length_scale}(a) and on a logarithmic scale in Fig.~\ref{fig:length_scale}(c). Again, we fit the data to the power-law
\begin{equation}
    \xi^{-1}\propto(1-T/T_c)^{n_\xi}\,.
    \label{eq:critical exponent of xi}
\end{equation}
We found $n_\xi=0.41\pm0.02$. The deviation from the linear regression at high temperatures may be a result of analysis failure since the penetration depth is no longer much smaller than the ring's height ($\lambda\;\cancel{\ll}\;h$). At low temperatures, we associate the deviation with heating caused by the strong current in the excitation coil, which cannot be accurately accounted for by the temperature calibration.

Alternative determination of $\xi$ is from $H_{c2}$~\cite{Tinkham2004IntroductionSuperconductivity} according to the equation
\begin{equation}
    \mu_0H_{c2}=\frac{\Phi_0}{2\pi\xi^2(T)}\,.
    \label{eq:Hc2 and xi}
\end{equation}
$\xi$ determined from $H_{c2}$ is presented on a linear scale with black squares in Fig.~\ref{fig:length_scale}(a) and $1/\xi$ on a logarithmic scale in panel (c) of the same figure for comparison. Here we also fit the data according to Eq.~\ref{eq:critical exponent of xi} and obtained $n_\xi=0.60\pm0.03$.

For further comparison, low-temperature measurements of $1/\xi$ from other methods have been added to Fig.~\ref{fig:length_scale}(c) (stars shaped): Resistivity \cite{Shruti2015AnisotropyFeTe0.55Se0.45}, ARPES \cite{Chiu2020ScalableSuperconductors}, and STM \cite{Wang2018EvidenceSuperconductor}. The resistivity measurement is, in fact, an $H_{c2}$ measurement, and the result obtained is close to the one obtained by the magnetization method ($\xi_{H_{c2}}/\xi_\text{Res}=1.6$ at $T=0$). The ARPES value $\xi_0$ is related to the GL $\xi$ at $T=0$, $\xi(0)$, by a factor of 0.74 \cite{Tinkham2004IntroductionSuperconductivity} (equation~4.24). The same factor was taken into account when converting the STM result. Unlike the $H_{c2}$ measurements, the results from the other two methods are closer to the linear regression of the Stiffnessometer method ($\xi_\text{Stiff}/\xi_\text{ARPES}=1.7$ and $\xi_\text{Stiff}/\xi_\text{STM}=1.9$ at $T\rightarrow 0$).
\section{\label{sec:knee data} Reproducibility and origin of the Knee}

\begin{figure}[h]
    \centering
        \includegraphics[width=\linewidth]{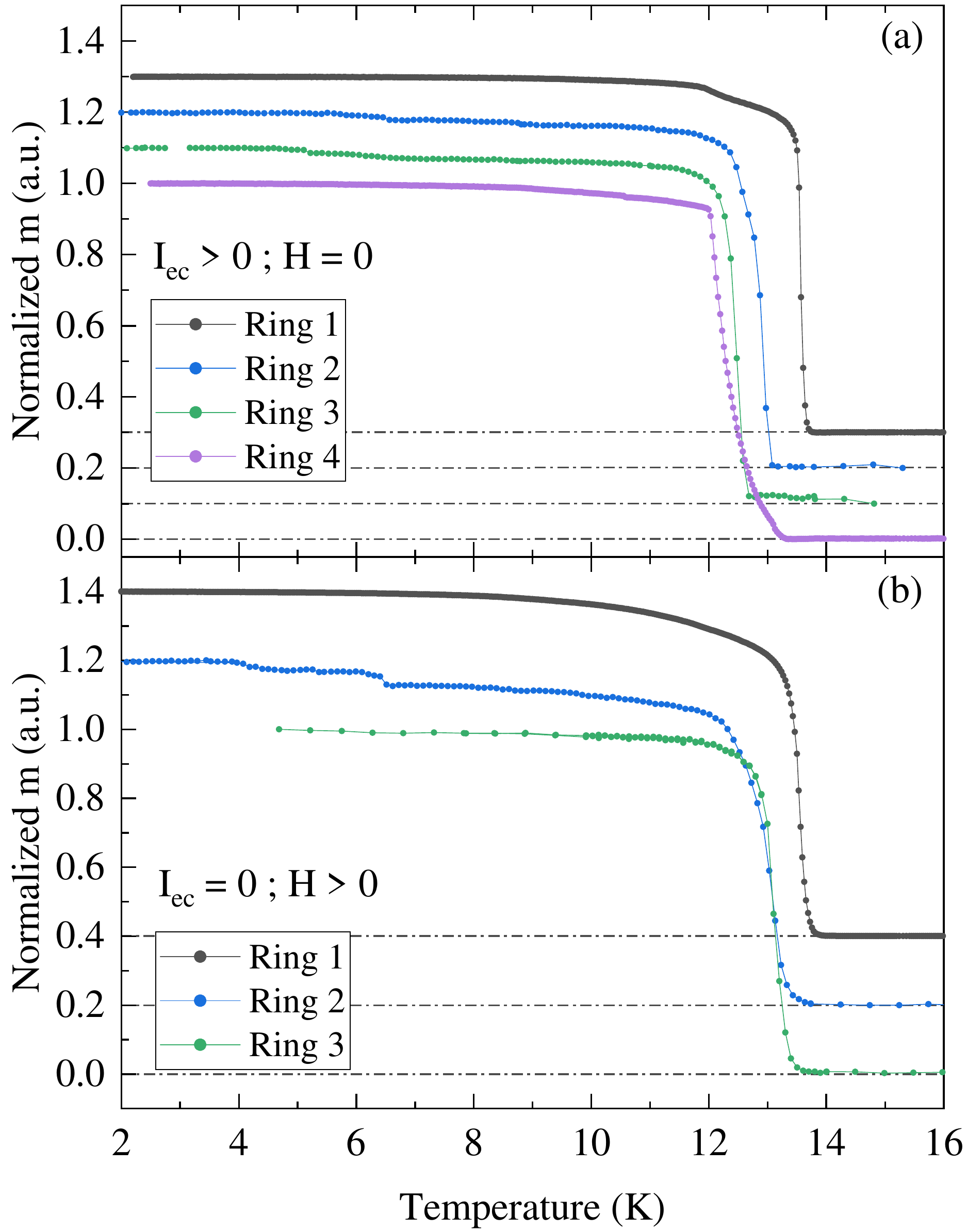}
    \caption{{\bf Reproducibility.} Normalized magnetic moment $m/m(T\rightarrow 0)$ vs. temperature for different rings. (a) in the presence of current in the excitation coil, as described in Sec.~\ref{subsec:stiffness}. (b) in the presence of an applied field perpendicular to the ring. The central ring of this research is 1. An offset is added for clarity.}
    \label{fig:KneeRingsComparison}
\end{figure}

To examine reproducibility we investigated more than one ring cut from different crystals of FST. A comparison between different rings from other crystals appears in Fig.~\ref{fig:KneeRingsComparison}. The figure shows the normalized, and shifted (for clarity), SC's moment as a function of the temperature in two cases: Panel (a) with a current in the EC and zero applied field, and panel (b) with an applied field and no EC current. The applied currents and fields are in the range $[5,10]$~mA and $[0.1,3]$~mT, respectively, but not necessarily equal for different rings. The main ring of this research is 1. The knee temperature and sharpness vary from ring to ring (Fig.~\ref{fig:KneeRingsComparison}(a)). Interestingly, multiple knees appear in the standard, in-field, measurement of ring~2 in panel (b).

\begin{figure}[h]
    \centering
        \includegraphics[width=\linewidth]{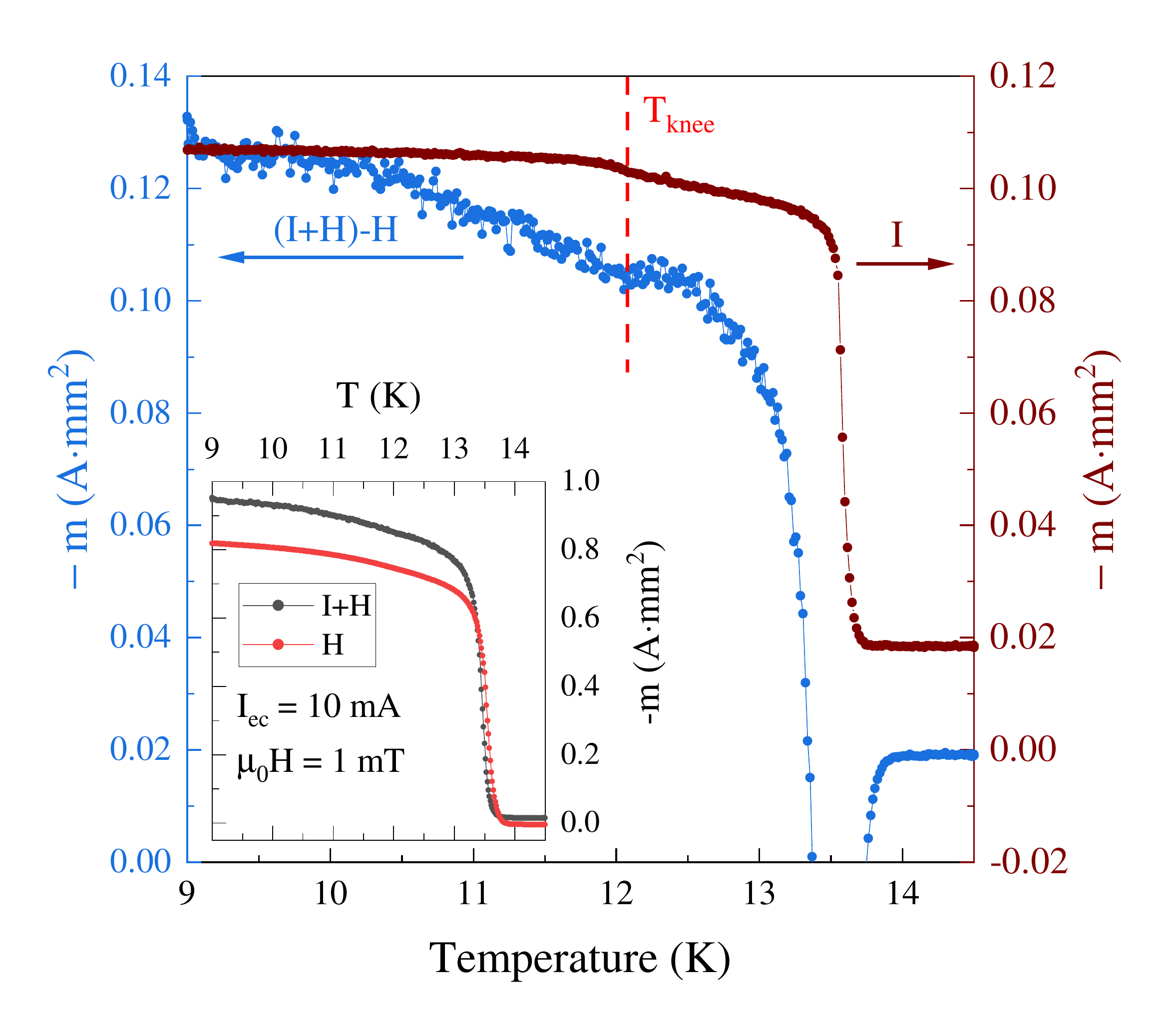}
    \caption{{\bf The knee's field dependence.} Temperature dependence of the difference between the ring's moment measured with excitation coil current and field, to a measurement with field only (blue-circles left Y-axis). The brown-circles right Y-axis is the measurement in zero field but current in the excitation coil. The data is shifted for clarity. The knee is at the same temperature regardless of the field. The inset shows the ring's moment vs. the temperature in the presence and absence of an applied field (1~mT) and current in the EC (10~mA). The data in the main panel is obtained by subtracting the two data sets in the inset.}
    \label{fig:kneeManipulation}
\end{figure}

It could be that the knee originates from the interaction of the superconducting order parameter with the underline ferromagnet. To test this possibility we measure the SC's magnetic moment vs. temperature in the presence of an applied field (ZFC) in the direction of the EC, with and without current in the coil. The raw data is shown in the inset of Fig.~\ref{fig:kneeManipulation}. The difference between the two measurements is presented in Fig.~\ref{fig:kneeManipulation}. For comparison, the measurement with current only is also displayed. The knee appears at the same temperature with and without the field.

\section{\label{sec: discussion} DISCUSSION}

We discuss the two major observations of this work, the knee, and the critical exponents, and examine the relation between measured quantities.

\subsection{\label{subsec: Knee Discussion} The Knee}

Mukasa {\it et al.}~[\onlinecite{Mukasa2021High-pressureSuperconductivity}]
present the nematic transition temperatures $T_s$ as a function of the Tellurium composition in Fe$_{1+y}$Se$_x$Te$_{1-x}$. The lowest temperature measured by x-ray diffraction is $13.3$~K, and there are no measurements close to the mid point $x=0.5$. Nevertheless, extrapolation of their data suggests that $T_s$ and $T_c$ cross each other near $x=0.5$ and that $T_s$ might drop below $T_c$. Perhaps nematic order is the origin of the knee. Alternatively, Peng Zhang {\it et al.}~[\onlinecite{Zhang2018ObservationSuperconductor}], suggest the existence of surface superconductivity in FST. We speculate that this might lead to two different SC $T_c$s, one for the bulk and one for the surface. FST is also known to have multiple Fermi surfaces. It could be that the knee is a result of the different temperature dependence of the SC order parameters on different bands.

Finally, there is always the possibility that the knee is a result of the geometrical imperfection of the ring. Such imperfections are difficult to account for in numerical simulations.

\subsection{\label{subsec:comparison with theory} Critical Exponents}

GL theory assumes, and BCS theory predicts, a linear temperature dependence of $\psi^2$. According to Eq.~\ref{eq:stiffness} this leads to the prediction that $n_{\rho}=1$. Our finding is not exactly as expected, but it is closer to unity than the results of $\mu$SR added to Fig.~\ref{fig:length_scale}(b). It should be pointed out that the $\mu$SR measurements are done in a fixed magnetic field, which becomes higher than $H_{c2}$ as one approaches $T_c$. The discrepancy between techniques could also result from an interaction between the applied field and the underline ferromagnet, as mentioned before. Similarly, standard GL predicts $n_\xi=0.5$. In this case, $\xi$ determined by the Stiffnessometer and $H_{c2}$ are equally far from the expected value.

If we relax the linear assumption, the GL theory also predicts $n_\rho/n_\xi=2$. We find
\begin{equation}
    n_\rho/n_\xi=2.22\pm0.12\,.
\end{equation}
The result obtained from the $\mu$SR and $H_{c2}$ methods gives $n_\rho^{\mu\text{SR}}/n_\xi^{H_{c2}}=0.88\pm0.08$, far from the GL expected value.

\subsection{\label{subsec: First critical field} First critical field}

The first critical field, $H_{c1}$, is related to $\lambda$ and $\xi$ \cite{DeGennes1999SuperconductivityAlloys} via 
\begin{equation}
    \mu_0H_{c1}=\frac{\Phi_0}{4\pi\lambda^2}\ln{\frac{\lambda}{\xi}}\,.
    \label{eq: Hc1 to k}
\end{equation}
An attempt to test this equation fails severely regardless of the experimental method used to determine the different quantities. Bendele {\it et al.}~\cite{Bendele2010AnisotropicTe0.5} addressed this problem by considering the demagnetization factor $D$. They introduced the equation
\begin{equation}
    B=\mu_{0}\left(m/V+H_\text{int}\right)\,,
\end{equation}
where $H_\text{int}=H_\text{ext}-D\cdot m/V$, $H_\text{int}$ and $H_\text{ext}$ are the internal and externally applied field, respectively, $\mu_0 H_{c1} \rightarrow B$ in Eq.~\ref{eq: Hc1 to k}. This calculation is very sensitive to the ring's volume and $D$ accuracy. In Sec.~\ref{subsec:Tc} we considered two options for $D$. If we adopt the disk option we get a much smaller $B$ than measured. If we consider the ring option we find a negative $B$ value. Sometimes additional constant is considered in Eq.~\ref{eq: Hc1 to k} that includes the effect of the hard core of the vortex line \cite{Bendele2010AnisotropicTe0.5,Sajilesh2018SuperconductingLaPtGe,Shang2019EnhancedSuperconductor}, but in our case, this effect is negligible. Once again we speculate that the failure of Eq.~\ref{eq: Hc1 to k} is a result of the underlining  ferromagnetism in FST.

\section{\label{sec:Conclusions} Conclusions}

We developed a method, ideal for magnetic superconductors close to $T_c$, to measure both the penetration depth $\lambda$ and coherence length $\xi$. For FST we find that $\lambda$ and $\xi$ are longer than previously reported and their temperature dependence agrees better with the GL predictions. A second transition, that looks like a knee, is observed at a temperature below $T_c$ in the stiffness measurements. Further experiments are required to determine whether this transition is due to either nematic order, surface superconductivity, multiple Fermi surfaces, or a simple geometrical effect.

\section{Acknowledgements}
We thank Amit Kanigel and Avior Almoalem for the sample.
We are grateful to the nano-center at Tel-Aviv university for the use of their femtosecond laser cutter. This research was supported by Israel Science Foundation personal grant number 3875/21 and the Nevet grant, Russel Berrie Nanotechnology Institute, Technion.



%

\begin{appendices}

\begin{figure}[H]
    \centering
    \includegraphics[width=\linewidth]{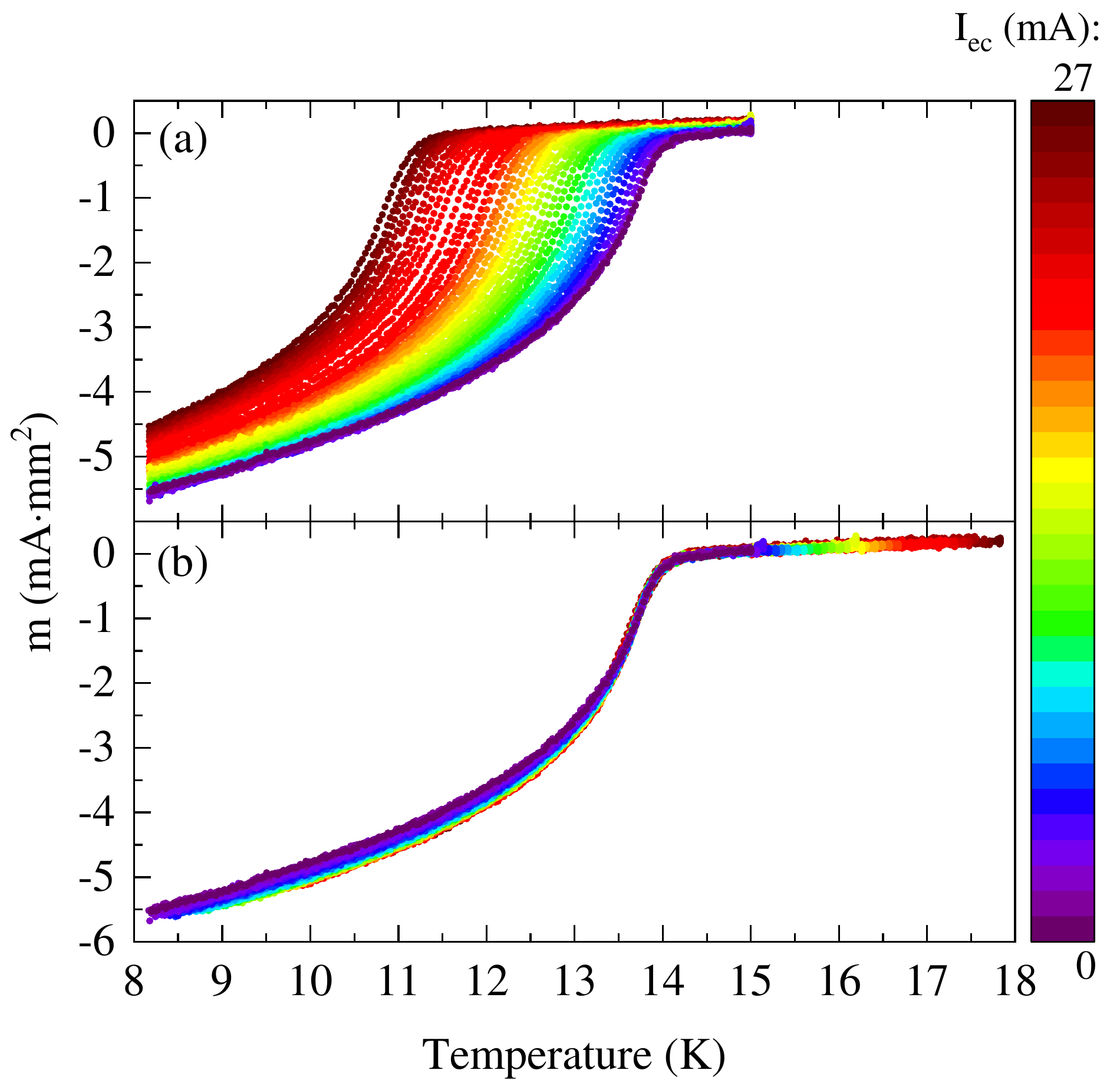}
    \caption{{\bf Temperature Calibration. } Temperature dependence of the magnetic moment of a disconnected FST ring in the presence of a magnetic field, repeated for different $I_\text{ec}$, as indicated by the colors. (a) Before the calibration. (b) After calibration.}
    \label{fig:1}
\end{figure}

\section{Temperature calibration}
\label{Apx:T_calibration}

\begin{figure}[H]
    \centering
    \includegraphics[width=\linewidth]{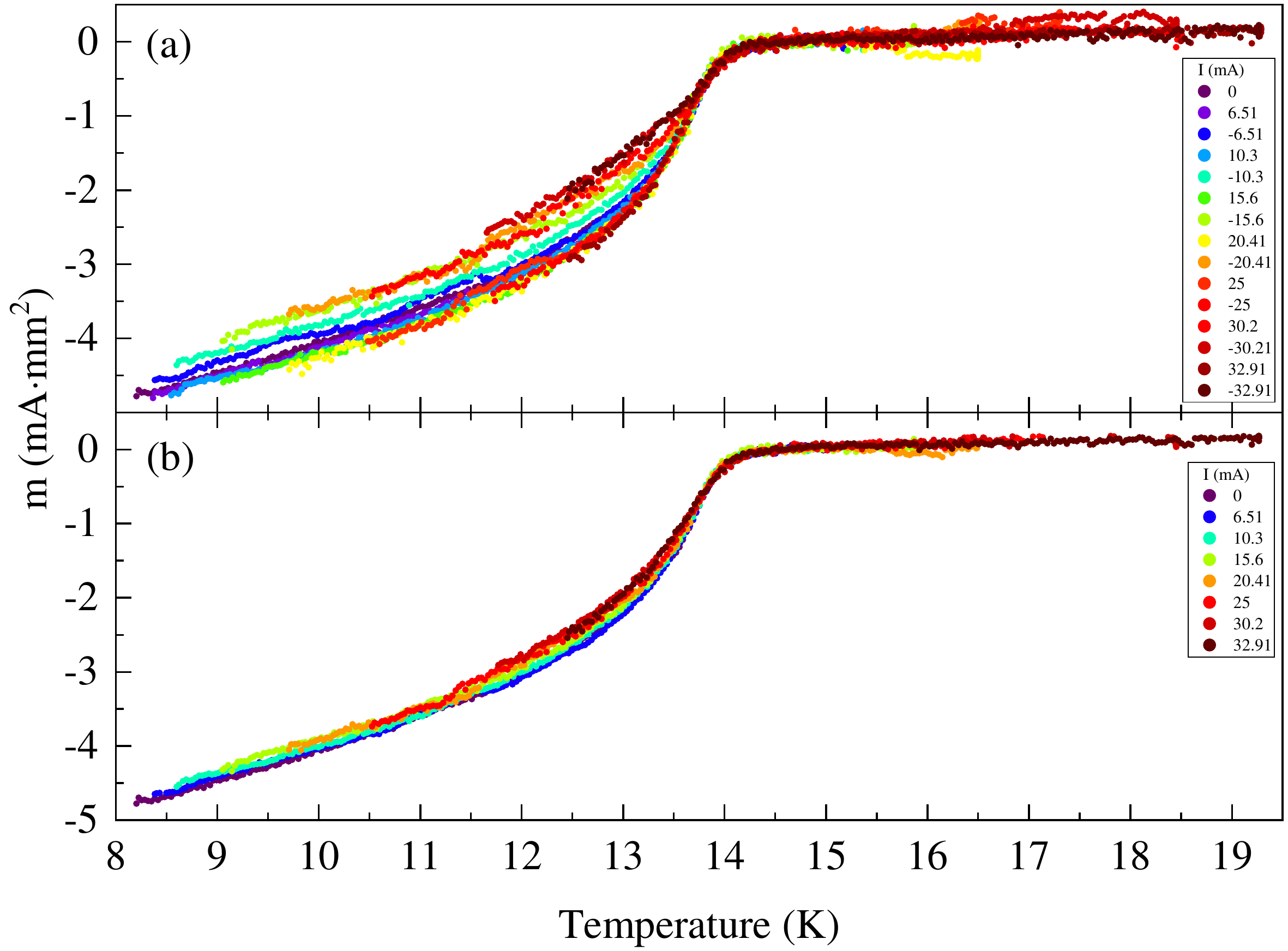}
    \caption{{\bf The influence of the leaking field from the excitation coil on the measurements.} (a) Calibrated measurements in the presence of positive and negative current values, as indicated by the colors.. (b) Averaging over the directions of the currents in (a).}
    \label{fig:field_leak}
\end{figure}

Due to the heat produced by the current in the EC, a temperature gradient is developed between the ring and the thermometer, s.t., the actual temperature of the sample $T$, and the temperature recorded by the chamber thermometer $T_\text{ch}$ are not the same. Our goal is to determine the sample temperature $T$ corresponding to each critical current $I_\text{ec}^c$ based on the chamber temperature $T_\text{ch}$.

\begin{figure}[h]
    \centering
    \includegraphics[width=\linewidth]{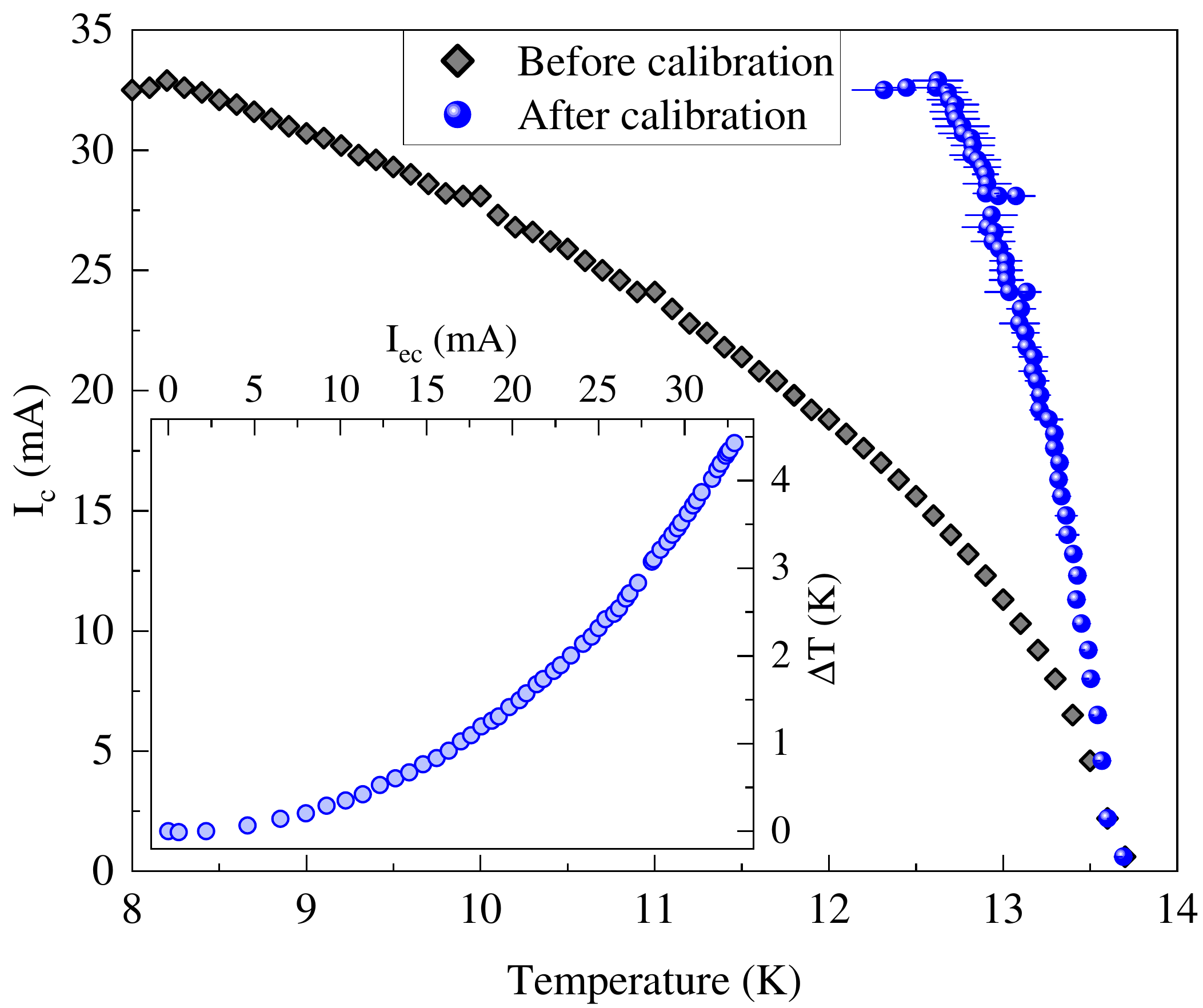}
    \caption{{\bf Critical currents before and after the calibration.} Critical current vs. the temperature before the calibration in gray-diamonds and after in blue-circles. The inset shows the temperature correction $\Delta T$ vs. the current in the excitation coil. The relation is approximately parabolic.}
    \label{fig:before&after_calibration}
\end{figure}

The calibration process is done by measuring the temperature dependence of the magnetic moment in the presence of an applied field of $\mu_0H\approx1$~mT (ZFC), similarly to Sec.~\ref{subsec:Tc}. However, this time we use a disconnected FST ring and repeat the measurement for different $I_\text{ec}$ values. The critical current values from Fig.~\ref{fig:MvsI_and_dMdIvsT}(b)-inset have been chosen to improve the accuracy. 

The current in the EC heats the sample but cannot generate a persistent current in the ring due to the disconnection. Nevertheless, there are two additional contributions of the EC current to the signal, and both are consequences of its finite length. A good way to understand them is from the EC signal in Fig.~2-inset and Fig.~3 in Ref.~\cite{Mangel2020Stiffnessometer:Application}: I) The second-order gradiometer is insensitive to any field uniform in space, but even around its center, the EC signal is not totally uniform mostly due to asymmetry of the coil (e.g., wires enter the coil from one side only). This contribution is identified from the measurement above $T_c$ and subtracted. The measurement results after this subtraction appear in Fig.~\ref{fig:1}(a); II) A field leakage from the EC, altering the field in the sample and the sample's moment accordingly. This field leakage could be partially canceled by measuring the moment in two current directions, as presented in Fig.~\ref{fig:field_leak}(a). The difference between measurements increases with the current while the zero current measurement stays in the middle. Averaging over both directions reduces the deviations due to field leakage, as in Fig.~\ref{fig:field_leak}(b). Notably, the magnitude of the field leaking from the coil at 10~mA current is estimated to be 0.03~mT.

Once these contributions are eliminated, we search for the temperature correction, $\Delta T$, for which $m(I_\text{ec},T_\text{ch}+\Delta T)$ collapses onto the one without the current $m(0,T)$ at the steepest part of the measurement's slope as seen in Fig.~\ref{fig:1}(b). The collapse is best close to $T_c$, but the correction is suitable for a wide range of temperatures. The relation generated between $I_\text{ec}$ and $\Delta T$ is given in Fig.~\ref{fig:before&after_calibration}-inset. 

\begin{figure}[H]
    \centering
    \includegraphics[width=\linewidth]{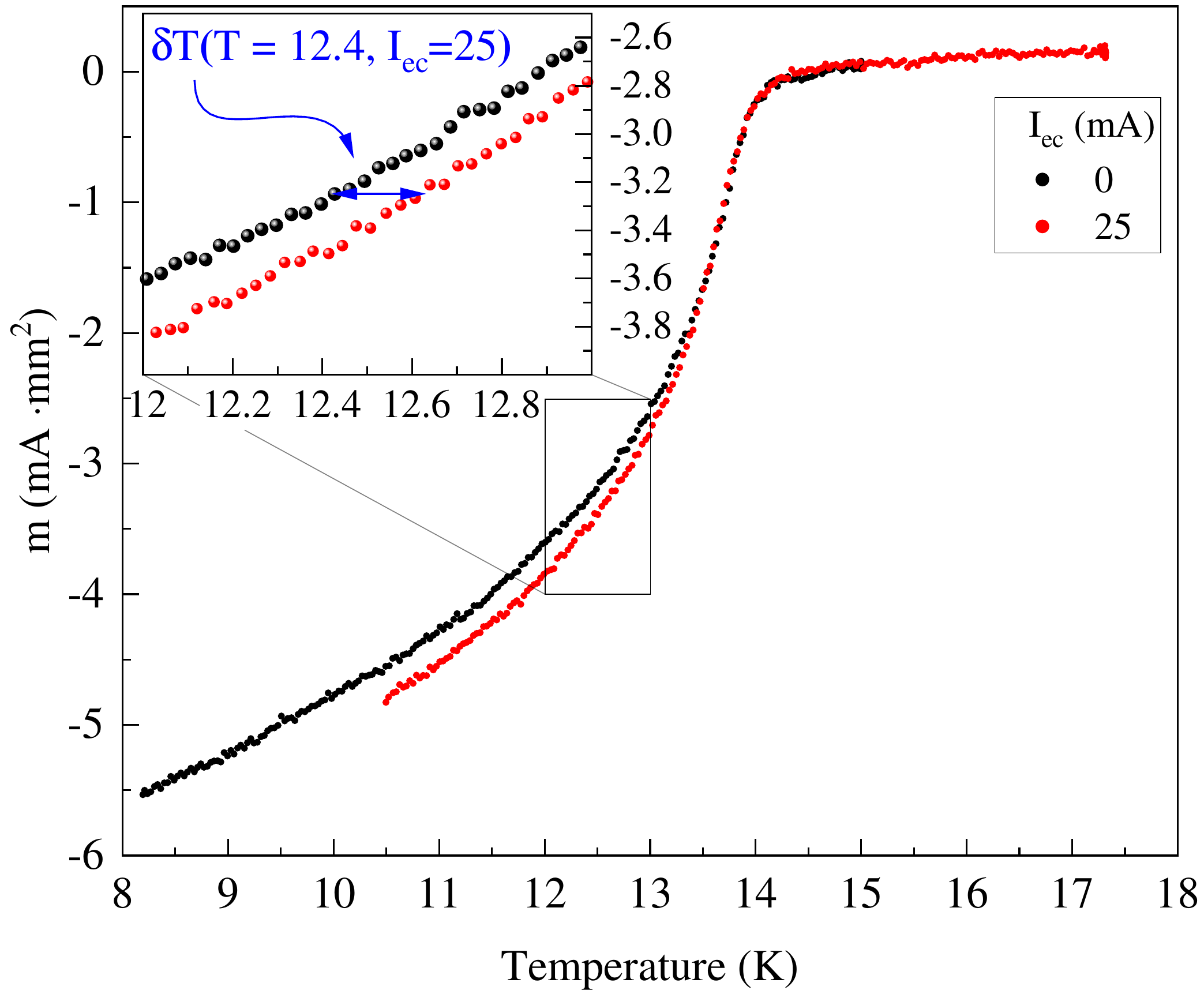}
    \caption{{\bf Estimation of the temperature calibration process errors}. Measurements after temperature calibration (from Fig.~\ref{fig:1}(b)) without current in the excitation coil in black-circles and with a current of $I_\text{ec}=25$~mA in red-circles. The error is estimated by the temperature difference between two points with the same moment value. It is represented by $\delta T$ and depends on the current and the temperature.}
    \label{fig:3}
\end{figure}

After the temperature correction, for each current, is set, we compare the measurement with the current to the one without. The error in the temperature correction is estimated by the temperature difference between points with the same moment value from both measurements. An example appears in Fig.~\ref{fig:3}. The errors depend on the current in the coil and temperature. Finally, in blue circles in Fig.~\ref{fig:before&after_calibration}, we present the SC critical current $I_\text{ec}^c$ as a function of the calibrated temperature $T$ with error-bars.\\

\end{appendices}

\end{document}